\documentclass[a4paper,12pt]{article}
\usepackage[english]{babel}
\usepackage[utf8]{inputenc}
\usepackage{cite}
\usepackage{amsthm,amsfonts,amsmath,mathrsfs}
\usepackage{slashed}
\usepackage{makeidx}
\usepackage{graphicx}
\usepackage{caption}

\usepackage{hyperref}
\usepackage{verbatim}

\newcommand{\be}{\begin{eqnarray}}
	\newcommand{\ee}{\end{eqnarray}}

\begin{document}
	
	\title{Ambiguities in the generation of CFJ-terms in a QED with dimension-5 operators in one loop} %On the one loop generation of CFJ-terms in 5-dimensional CPT-odd terms in an extended QED}
	
	\maketitle
	\begin{center}
	{     H. G. Fargnoli$^a$,	
              J. C. C. Felipe$^{b}$,
		   G. Gazzola$^c$
	}\\[2em]
	{\sl ${}^a$ Departamento de Matemática e Matemática Aplicada, Instituto de Ciências Exatas e Tecnológicas, Universidade Federal de Lavras, Caixa Postal 3037, 37200-900, Lavras, MG, Brasil}\\
	{\sl ${}^b$ Departamento de Estat\' istica, F\'isica e Matem\'atica, Universidade Federal de São João del Rei, Rod. MG 443, Km 7, 36497-899 - Ouro Branco-MG, Brasil}\\
	{\sl ${}^c$Instituto de Engenharia, Ci\^encia e Tecnologia, Universidade Federal dos Vales do Jequitinhonha e Mucuri, Avenida Um, 4050 - 39447-790 -Cidade Universit\'aria -Jana\'uba-MG, Brasil}
	\end{center}
	
	\begin{abstract}
	\noindent	%In this paper, we consider a Lorentz symmetry breaking Electrodynamics (QED) with 5-dimensional CPT-odd terms in one loop. Applying an independent regularization method, we split the divergences into dependent (surface terms) and independent terms of the regularization scheme. We have shown how such surface terms, which can contribute to the generation of CFJ terms, remain undetermined even when evoking Ward-Takahashi identity. Our results align with previous findings in the literature while highlighting the importance of a consistent treatment of regularization ambiguities.		
    In this paper, we investigate a Lorentz symmetry breaking extension of Quantum Electrodynamics (QED), incorporating 5-dimensional CPT-odd terms at the one-loop level. Employing an independent regularization method, we systematically separate the divergences into regularization-independent contributions and regularization-dependent (surface terms). We demonstrate that these surface terms, potentially contributing to the radiative generation of Carroll-Field-Jackiw (CFJ) terms, remain undetermined even when imposing the Ward-Takahashi identity. Our results are consistent with existing literature and emphasize the necessity of a careful and consistent treatment of regularization ambiguities in quantum field theories involving symmetry breaking extensions.
	\end{abstract}
	
	\section{Introduction}
	
	In the last decades, theories that violate Lorentz and CPT symmetries have aroused great interest from researchers and have been extensively studied. After the consolidation of the standard model extension (SME) \cite{Colladay1997,Colladay1998}, many theoretical and experimental aspects have been addressed (see \cite{Kost2011} for a list of experimental results and bounds to the SME parameters). More recently, \cite{KosteleckyPRD2019} has expanded the SME, adding gauge-invariant interactions at arbitrary mass dimensions.% $d$.
	
	Usually, Lorentz-violating terms are introduced in quantum field theories as quantities proportional to constant vectors or tensors \cite{PetrovEPJC2021, Colladay1997, Colladay1998, Kost2004}. These terms can generate well-known minimal couplings in the physical dimension ($d=4$) \cite{Kost2002}. In this sense, perturbative aspects within Lorentz-violating models become essential if we consider the Carroll-Field-Jackiw (CFJ) term, generated by including a Lorentz-violating term in the Lagrangian of the model \cite{Carroll1990, Jackiw1999}.
	
	In this paper, we revisit the possibility of generating CFJ terms in one-loop perturbative order in a Lorentz-violating extended quantum electrodynamics (QED). We considered non-minimal 5-dimensional CPT-odd terms. In \cite{PetrovEPJC2021}, the authors showed that the generation of a CFJ term is intrinsically related to regularization schemes used in the intermediate steps of the calculation, and they obtained different results using different regularization prescriptions.
	
	Generally, when one uses perturbation theory in quantum field theories, it is necessary to implement a regularization method to deal with divergences. However, such methods can spoil crucial aspects of the theory, like gauge symmetry, or they can introduce ambiguities throughout calculations. In these cases, a regularization-independent framework can split the divergences into regularization-dependent (surface terms) and regularization-independent terms. Then, one can fix undetermined parameters on symmetry or phenomenological grounds. For instance, in \cite{GazzolaEPL2013}, the authors have fixed an ambiguity using a symmetry of the theory. On the other hand, in \cite{Gazzola2012}, the symmetry of the theory was not enough to fix the ambiguities.
	 
	Therefore, we proceed to apply a regularization independent scheme, namely, Implicit Regularization (IR) \cite{Battistel1998,Ferreira2012,FelippeEPJC2022}, to shed light on this issue in the presented theory. Our calculations have explicitly separated the intrinsic divergences (regularization-independent terms) and the surface terms (regularization-dependent terms), which have not been done so far in the present model. We have shown how such surface terms show up in the possible generation of CFJ terms.
	
	This paper is organized as follows. In Section \ref{model}, we present the model, the relevant amplitudes to perform the calculations, and the basic features of IR. In Section \ref{results}, we show the results for each amplitude at one-loop level and evoke Ward-Takahashi identify in an attempt to fix the surface terms. Section \ref{conclusion} is dedicated to concluding remarks.
		
	\section{The extendend Lorentz-violating QED model}\label{model}
	
	In \cite{KosteleckyPRD2019}, the authors construct a general (with arbitrary mass dimension) gauge-invariant theory with Lorentz and CPT violation. In particular, in Table I, \cite{KosteleckyPRD2019} displays the terms of mass dimension 5 in the fermion Lagrangian $\mathcal{L}_{\psi}$. In this paper, we are interested in generating a CFJ term perturbatively in one loop. So we need to consider only odd-rank tensors in four-dimensional spacetime, and our Lagrangian reduces to \cite{PetrovEPJC2021}:
	
	\begin{eqnarray}\label{Lagrangiana}
		\mathcal{L}_{\psi} &=& \bar{\psi}\left[i{\slashed \partial} + \left(a^{(5)\mu\alpha\beta}+b^{(5)\mu\alpha\beta}\gamma_5\right)\gamma_{\mu}\partial_{\alpha}\partial_{\beta}-m-e{\slashed A}\right.\nonumber\\ && +  ie\left(a^{(5)\mu\alpha\beta}+b^{(5)\mu\alpha\beta}\gamma_5\right)\gamma_{\mu}\left(2A_{\beta}\partial_{\alpha}+\partial_{\alpha}A_{\beta}\right) \\
		&& \left.  -2e\left(a_{F}^{(5)\mu\alpha\beta}+b_{F}^{(5)\mu\alpha\beta}\gamma_5\right)\gamma_{\mu}\partial_{\alpha}A_{\beta}-e^2\left(a^{(5)\mu\alpha\beta}+b^{(5)\mu\alpha\beta}\gamma_5\right)\gamma_{\mu}A_{\alpha}A_{\beta}\right]\psi,\nonumber
	\end{eqnarray}
where the $a$'s and $b$'s are the Lorentz-violating coefficients.
	
	Considering the one-loop effective action of the gauge field and the contributions of first order in the Lorentz-violating coefficients and second order in $A^{\mu}$, we focus on \cite{PetrovEPJC2021}:
	
	\begin{equation}
		S_{\rm eff}^{(1,2)} = \frac{i}{2}\int{ d^4 x \sum_{n=1}^6 \Pi^{\mu\nu}_n A_\mu A_\nu } \label{Seff}
	\end{equation}
    where $S_{\rm eff}^{(1,2)}$ is the power series up to second order of the effective action \cite{PetrovEPJC2021}, with
	\begin{eqnarray}
		\Pi^{\mu\nu}_1 &=& \displaystyle 2e^2\int \frac{d^4p}{(2\pi)^4} \mbox{tr}\ S(p)\left(a^{(5)\lambda\mu\nu}+b^{(5)\lambda\mu\nu}\gamma_5\right)\gamma_{\lambda},\\
		\Pi^{\mu\nu}_2 &=& \displaystyle - e^2\int \frac{d^4p}{(2\pi)^4} \mbox{tr}\ S(p)\left(a^{(5)\alpha\beta\mu}+b^{(5)\alpha\beta\mu}\gamma_5\right)\gamma_{\alpha}\left(2p_{\beta}+k_{\beta}\right)S(p-k)\gamma^{\nu}, \label{DefPi2} \\
		\Pi^{\mu\nu}_3 &=& \displaystyle - e^2\int \frac{d^4p}{(2\pi)^4} \mbox{tr}\ S(p)\gamma^{\mu}S(p-k)\left(a^{(5)\alpha\beta\nu}+b^{(5)\alpha\beta\nu}\gamma_5\right)\gamma_{\alpha}\left(2p_{\beta}-3k_{\beta}\right),\label{DefPi3}\\
		\Pi^{\mu\nu}_4 &=& \displaystyle e^2\int \frac{d^4p}{(2\pi)^4} \mbox{tr}\left[S(p)\left(a^{(5)\alpha\beta\lambda}+b^{(5)\alpha\beta\lambda}\gamma_5\right)\gamma_{\alpha}p_{\beta}p_{\lambda}S(p)\gamma^{\mu}S(p-k)\gamma^{\nu}\right. \label{DefPi4} \\
		& &\left.+ S(p)\gamma^{\mu}S(p-k)\left(a^{(5)\alpha\beta\lambda}+b^{(5)\alpha\beta\lambda}\gamma_5\right)\gamma_{\alpha}(p-k)_{\beta} (p-k)_{\lambda}S(p-k)\gamma^{\nu}\right],\nonumber \\
		\Pi^{\mu\nu}_5 &=& \displaystyle -2ie^2\int \frac{d^4p}{(2\pi)^4} \mbox{tr}\  S(p)\left(a_{F}^{(5)\alpha\beta\mu}+b_{F}^{(5)\alpha\beta\mu}\gamma_5\right)\gamma_{\alpha}k_{\beta}S(p-k)\gamma^{\nu},\\
		\Pi^{\mu\nu}_6 &=& \displaystyle -2ie^2\int \frac{d^4p}{(2\pi)^4}\mbox{tr}\ S(p)\gamma^{\mu}S(p-k)\left(a_{F}^{(5)\alpha\beta\nu}+b_{F}^{(5)\alpha\beta\nu}\gamma_5\right)\gamma_{\alpha}\left(-k_{\beta}\right),
	\end{eqnarray}
	
	\noindent and $S(p)=\dfrac{1}{{\slashed p}+m}$.
	
	To avoid ambiguities (or to make them explicit), we will perform the above calculations\footnote{Our results were obtained using slightly different methods of IR. One method is described in \cite{Battistel1998,Ferreira2012} and the other in \cite{FelippeEPJC2022}. The results were identical as expected.} in the Implicit Regularization framework \cite{Battistel1998,Ferreira2012}. 
	
	\subsection{Implicit regularization in a nutshell}
	
	An important feature of Implicit Regularization is that the dimension of the theory is not modified which makes it adequate to study theories with objects like $\gamma_5$ matrices and Levi-Civita symbols. Implicit Regularization also allows isolation of possible ambiguities in well-defined objects which can be further investigated.
	
	This method has been applied successfully in several situations. For instance, in the study of a semi-relativistic model for the graphene \cite{GazzolaEPL2013}, the method was able to reveal an ambiguity which had plagued the perturbative computation of the model for the graphene conductivity. This result made possible the mapping for each different result which had been published so far, and using the underlying symmetry of the model the authors fixed this ambiguity. The method also was used in the study of gravitational effective theories \cite{Felipe2011}, where gravitational contributions to asymptotic freedom of electrodynamics were analyzed. In this case, some authors found that quadratic divergences made the Electrodynamics of Maxwell asymptotically free when gravitational effects are taken into account. However in \cite{Felipe2011} it was shown that these contributions were ambiguous. A last example, the method also was applied in the study of gravitational anomalies in two dimensions \cite{Dallabona2024}. For a recent review of Implicit Regularization see \cite{Perdomo2021} (see also references therein). A comparison between Implicit Regularization and other regularizations methods (including Dimensional Regularization) was done in \cite{Gnendiger2017} (see also \cite{Cherchiglia2021}).
	
	Within Implicit Regularization framework, it is assumed that the integrals are regularized by some implicit regulator. Then it is used the identity:
	\begin{equation}
	\frac{1}{[(k_{i} \pm p )^{2}-m^{2}]}= \sum_{j=0}^{N}\frac{(-1)^{j}(k_{i}^{2} \pm 2k_{i}\cdot p)^{j}}{(p^{2} - m^{2})^{j+1}} + \frac{(-1)^{N+1}(k_{i}^{2} \pm 2k_{i}\cdot p)^{N+1}}{(p^{2}-m^{2})^{N+1}[(k \pm p)^{2}-m^{2}]},
	\end{equation}
	where $k_i$ are external momenta, $p$ is an internal momentum and $N$ is such that the last term is finite. The value of $N$ depends on the theory degree of divergence. It is important to notice that this identity expands the divergence of a term into its subset of divergencies, until one starts to get finite terms. For such process, and to be able to display properly each divergent integral, one defines the basic divergent integrals as
	\begin{eqnarray}
		\displaystyle I_{\text{quad}}(m^2) &=& \int\frac{d^4p}{(2\pi)^4} \frac{1}{p^2-m^2} \label{defIquad}\\
		\displaystyle I_{\text{log}}(m^2) &=& \int\frac{d^4p}{(2\pi)^4} \frac{1}{(p^2-m^2)^2},\label{defIlog}
	\end{eqnarray}
	and so on, which are not calculated explicitly. We should stress that these integrals do not depend on external momenta. Besides the finite parts, we can also get surface terms like:
	\begin{align}
		\displaystyle &c_1 g_{\mu\nu} \equiv \frac{1}{4}\ g_{\mu\nu} \int\frac{d^4p}{(2\pi)^4} \frac{1}{(p^2-m^2)^2} - \int\frac{d^4p}{(2\pi)^4} \frac{p_{\mu}p_{\nu}}{(p^2-m^2)^3} && \label{surface_term} \\
        \displaystyle &c_2 g_{\mu\nu} \equiv \frac{1}{2}\ g_{\mu\nu} \int\frac{d^4p}{(2\pi)^4} \frac{1}{p^2-m^2} - \int\frac{d^4p}{(2\pi)^4} \frac{p_{\mu}p_{\nu}}{(p^2-m^2)^2} && \\
		\displaystyle &c_3 g_{\{\mu\nu}g_{\alpha\beta\}} \equiv \frac{1}{24}\ g_{\{\mu\nu}g_{\alpha\beta\}} \int\frac{d^4p}{(2\pi)^4} \frac{1}{(p^2-m^2)^2} - \int\frac{d^4p}{(2\pi)^4} \frac{p_{\mu}p_{\nu}p_{\alpha}p_{\beta}}{(p^2-m^2)^4} \\
        %
        %\displaystyle &c_3 \left(g_{\mu\nu}g_{\alpha\beta}+g_{\mu\alpha}g_{\nu\beta}+g_{\mu\beta}g_{\alpha\nu}\right) \equiv \frac{1}{24} \left(g_{\mu\nu}g_{\alpha\beta}+g_{\mu\alpha}g_{\nu\beta}+g_{\mu\beta}g_{\alpha\nu}\right) \times \\
        %&\hspace{5cm}\times \int\frac{d^4p}{(2\pi)^4} \frac{1}{(p^2-m^2)^2} - \int\frac{d^4p}{(2\pi)^4} \frac{p_{\mu}p_{\nu}p_{\alpha}p_{\beta}}{(p^2-m^2)^4} \nonumber \\
        %
        \displaystyle &c_4 g_{\{\mu\nu}g_{\alpha\beta\}} \equiv \frac{1}{8}\ g_{\{\mu\nu}g_{\alpha\beta\}}  \int\frac{d^4p}{(2\pi)^4}\frac{1}{p^2-m^2} - \int\frac{d^4p}{(2\pi)^4} \frac{p_{\mu}p_{\nu}p_{\alpha}p_{\beta}}{(p^2-m^2)^3}  \\
        %\displaystyle &c_4 \left(g_{\mu\nu}g_{\alpha\beta}+g_{\mu\alpha}g_{\nu\beta}+g_{\mu\beta}g_{\alpha\nu}\right) \equiv \frac{1}{8} \left(g_{\mu\nu}g_{\alpha\beta}+g_{\mu\alpha}g_{\nu\beta}+g_{\mu\beta}g_{\alpha\nu}\right)  \times  \\
        %&\hspace{5cm}\times \int\frac{d^4p}{(2\pi)^4}\frac{1}{p^2-m^2} - \int\frac{d^4p}{(2\pi)^4} \frac{p_{\mu}p_{\nu}p_{\alpha}p_{\beta}}{(p^2-m^2)^3} \nonumber \\
        \displaystyle &c_5 g_{\{\mu\nu} g_{\alpha \beta} g_{\sigma \delta\}} \equiv \frac{1}{192}\ g_{\{\mu\nu} g_{\alpha \beta} g_{\sigma \delta\}} \int\frac{d^4p}{(2\pi)^4} \frac{1}{(p^2-m^2)^2}  \\ &\hspace{3cm}- \int\frac{d^4p}{(2\pi)^4} \frac{p_{\mu}p_{\nu}p_{\alpha}p_{\beta}p_{\sigma}p_{\delta}}{(p^2-m^2)^5} \nonumber
	\end{align}
	and so on. In last equations, $g_{\{\mu\nu}g_{\alpha\beta\}}$ is the symmetric index permutation for the metric tensor: $g_{\mu\nu}g_{\alpha\beta}+g_{\mu\alpha}g_{\nu\beta}+g_{\mu\beta}g_{\nu\alpha}$, and similarly for the others. The surface terms depend on the regularization scheme. For instance, if one explicitly computes such terms using Dimensional Regularization (DR), their results are null.
    
    Here, a comment is in order. In general, there are limitations on the use of symmetrical integration when considering contractions of internal momentum on Feynman integrals. Consider, for instance, the substitution $k^2 \rightarrow k_{\alpha}k_{\beta}g^{\alpha\beta}$ or the symmetric replacement $k_{\alpha}k_{\beta} \rightarrow \frac{1}{4}k^{2}g_{\alpha\beta}$. When applied to a simple integral (assuming it is regularized), we obtain
    \begin{align}
        \int \frac{d^4p}{(2\pi)^4}\frac{p^{2}}{p^{2}(p-k)^{2}}&=\int \frac{d^4p}{(2\pi)^4}\frac{1}{(p-k)^{2}} \nonumber\\  &=\lim_{\mu^{2}\rightarrow 0} \int \frac{d^4p}{(2\pi)^4}\frac{1}{(p-k)^{2}-\mu^{2}} \nonumber \\
        &=\lim_{\mu^{2}\rightarrow 0} \int \frac{d^4p}{(2\pi)^4}\frac{1}{p^{2}-\mu^{2}}=0.\label{nosym}
    \end{align}
    However, if symmetric integration is employed, we instead find
    \begin{equation}
 \int \frac{d^4p}{(2\pi)^4}\frac{g_{\alpha\beta}p^{\alpha}p^{\beta}}{p^{2}(p-k)^{2}}= \frac{-i}{96\pi^{2}}k^{2}.\label{sym}
    \end{equation}
    
    As we can see, there is an evident ambiguity in the result when we compare the results \eqref{nosym} and \eqref{sym}, respectively. This discrepancy highlights that the result of such integrals is sensitive to the choice of regularization prescription.  As we can see, the symmetric integration can not be used in a general way \cite{Perez-Victoria2001}, \cite{RosadoEPJC2025}. A concrete example of this limitation, relevant to the present context, can be found in one of the results obtained in \cite{PetrovEPJC2021}. %particularly in the context of integrals lacking a well-defined dimensional regularization prescription, as used on \cite{PetrovEPJC2021}.

    It is important to stress that within our results, using Implicit Regularization, no symmetric integration was used.
	
	\section{Results and discussion}\label{results}
	
	We begin with the simplest contribution. Because of parity, the first integral of the polarization tensor vanishes:
	\begin{equation}
		\Pi_1^{\mu\nu} = 0.
	\end{equation}

Secondly, we discuss the $b_F^{(5)\mu\alpha\beta}$ terms. Remarkably, these results are null since their integrals vanish individually in one-loop order (including the finite parts). That is:
    
    \begin{equation}
        \Pi_{5,b_F}^{\mu\nu} = \Pi_{6,b_F}^{\mu\nu} = 0 \label{resPi56bf}.
    \end{equation}
    
    We emphasize that since no surface term is left in these terms, this result is independent of the regularization scheme, \textit{i.e.}, there is no ambiguity whatsoever. 
    
    Before we discuss the other integrals, we should add that if we analyze the Lagrangian of the model, we can recognize that the tensors $a^{(5)\mu\alpha\beta}$ and $b^{(5)\mu\alpha\beta}$ are symmetric in the last two indices and the terms $a_F^{(5)\mu\alpha\beta}$ and $b_F^{(5)\mu\alpha\beta}$ are antisymmetric (under the exact change of indices). Therefore, for the sake of comparison, we used the same choices as the authors in \cite{PetrovEPJC2021} to rewrite some of the rank-3 tensors. We can write them as a combination of a rank-1 tensor multiplied by the metric (in the symmetric case) and the Levi-Civita tensor (in the antisymmetric case):
	
	\begin{align}
		a^{(5)\mu\alpha\beta} &= a^{\alpha}g^{\mu\beta} + a^\beta g^{\mu\alpha} \\ %tirei: a^\mu g^{\alpha\beta} \label{a_exp} 
		b^{(5)\mu\alpha\beta} &= b^\mu g^{\alpha\beta} + b^{\alpha}g^{\mu\beta} + b^\beta g^{\mu\alpha} \label{b_exp} \\
		a_F^{(5)\mu\alpha\beta} &= \epsilon_{\ \ \ \; \gamma}^{\mu\alpha\beta} a_F^\gamma \label{aF_exp}.
	\end{align}

    %%%para trabalhar aqui: mudar a ordem para deixar mais claro que esse resultado é bem geral. e não depende dessas definições acima.
	%Remarkably, it is not necessary to display a definition of the tensor $b_F^{(5)\mu\alpha\beta}$ since its integrals vanish in one-loop order (including the finite parts). That is:
    %\begin{equation}
	%	\Pi_{5,b_F}^{\mu\nu} = \Pi_{6,b_F}^{\mu\nu} = 0 \label{resPi56bf}.
	%\end{equation}

    %We emphasize that since no surface term is left in these terms, this result is independent of the regularization scheme.
    
    It is important to note that we separated the parts for the $a$ and $b$ tensors due to the presence of the $\gamma_5$ matrix, which modifies the traces and leads to very different computations for each integral.

    The remaining divergent parts for $\Pi_5^{\mu\nu}$ and $\Pi_6^{\mu\nu}$ were computed as:
    %\begin{align}
    %    \Pi_{5,a_F}^{\mu\nu} &= 	8 i e^2 a_F^\alpha k^\beta \varepsilon^{\mu\nu}_{\ \  \alpha\beta} \left\{ 		\left[ 4c_1 k^2 - 2 c_2 - 8k^2 c_3		\right] + \frac{1}{3} I_{\log}(m^2) k^2 \right\} \\
		%\Pi_{5,a_F}^{\mu\nu} &= 
		%8 i e^2 \left\{ a_F^\alpha k^\beta \varepsilon^{\mu\nu}_{\ \  \alpha\beta} 
		%\left[ 4c_1 k^2 - 2 c_2 - 8k^2 c_3
		%\right]
		%\right\} \ + \nonumber \\
        %& \ \ \ \ 
		%+\frac{8}{3} i e^2 I_{\log}(m^2) k^2 a_F^\alpha k^\beta \varepsilon^{\mu\nu}_{\ \  \alpha\beta}  \\
        %\Pi_{6,a_F}^{\mu\nu} &=		8 i e^2 a_F^\alpha k^\beta \varepsilon^{\mu\nu}_{\ \  \alpha\beta} \left\{ \left[ 4c_1 k^2 - 2 c_2 - 8k^2 c_3		\right]+\frac{1}{3} I_{\log}(m^2) k^2        \right\} 
        %\Pi_{6,a_F}^{\mu\nu} &=
		%8 i e^2 \left\{ a_F^\alpha k^\beta \varepsilon^{\mu\nu}_{\ \  \alpha\beta} \left[ 4c_1 k^2 - 2 c_2 - 8k^2 %c_3
		%\right]
		%\right\} + \nonumber \\
		%& \ \ \ \ +\frac{8}{3} i e^2 I_{\log}(m^2) k^2 a_F^\alpha k^\beta \varepsilon^{\mu\nu}_{\ \  \alpha\beta} 
	%\end{align}
    \begin{equation}
        \Pi_{5,a_F}^{\mu\nu} = \Pi_{6,a_F}^{\mu\nu} =
		8 i e^2 a_F^\alpha k^\beta \varepsilon^{\mu\nu}_{\ \  \alpha\beta} \left\{     4c_1 k^2 - 2 c_2 - 8k^2 c_3
    		+\frac{1}{3} I_{\log}(m^2) k^2
        \right\} .
    \end{equation}
    
    These results are the first indication that we may have the generation of CFJ terms since both the divergences and the surface terms are followed by a Levi-Civita tensor.
    
    This type of result is also seen in the computation of \eqref{DefPi2}, \eqref{DefPi3} and \eqref{DefPi4}:
    \begin{align}
		\Pi_{2,b}^{\mu\nu} &= 0 \\
		\Pi_{3,b}^{\mu\nu} &= 0 \\
		\Pi_{4,b}^{\mu\nu} &= 	8 i e^2 b^\alpha k^\beta \varepsilon^{\mu\nu}_{\ \  \alpha\beta}\ \bigg\{
		c_1(12m^2-5k^2)+3c_2+86k^2c_3-36c_4-384k^2c_5		 \nonumber \\
		& \hspace{3.5cm} \left. -\frac{1}{3} I_{\log}(m^2) (k^2+9m^2 ) 
		+3 I_{\rm quad}(m^2) \right\}.
        %\Pi_{4,b}^{\mu\nu} &= 	8 i e^2 \left\{ b^\alpha k^\beta \varepsilon^{\mu\nu}_{\ \  \alpha\beta} 
		%\left[c_1(12m^2-5k^2)+3c_2+86k^2c_3-36c_4-384k^2c_5
		%\right]
		%\right\} \nonumber \\
		%&\ -\frac{8}{3} i e^2 I_{\log}(m^2) (k^2+9m^2 )b^\alpha k^\beta \varepsilon^{\mu\nu}_{\ \  %\alpha\beta} 
		%+24 i e^2 I_{\rm quad}(m^2) b^\alpha k^\beta \varepsilon^{\mu\nu}_{\ \  \alpha\beta} 
	\end{align}

    Following the same prescription, the results for the remaining terms were computed as:
    \begin{align}
        \Pi_{2,a}^{\mu\nu} &= \frac{8}{3}  e^2  I_{\log}(m^2) a^\alpha k_\alpha (k^\mu k^\nu - k^2 g^{\mu\nu})  \\
            & \hspace{0.5cm}  -8e^2   \{a^\mu k^\nu[k^2(-5c_1+96c_3  -432c_5)+6c_2-28c_4] \nonumber \\
            & \hspace{1.8cm}  +a^\nu k^\mu[k^2(-c_1+12c_3-48c_5)+c_2-4c_4]  \nonumber \\
             & \hspace{1.8cm} + 2a^\alpha k_\alpha k^\mu k^\nu(c_1+4c_3-48c_5) \nonumber \\
             & \hspace{1.8cm} -2a^\alpha k_\alpha g^{\mu\nu}[k^2(c_1-16c_3+48c_5)+4c_4]\} \nonumber \\
		\Pi_{3,a}^{\mu\nu} &= \frac{8}{3}  e^2  I_{\log}(m^2) a^\alpha k_\alpha (k^2 g^{\mu\nu}- k^\mu k^\nu) \\
             & \hspace{0.5cm}-8e^2   \{a^\mu k^\nu[k^2(-c_1+12c_3 -48c_5)+c_2-4c_4] \nonumber \\
            & \hspace{1.8cm}  +a^\nu k^\mu[3k^2(-7c_1+48c_3-144c_5)+10c_2-28c_4] \nonumber \\
            & \hspace{1.8cm}  +2a^\alpha k_\alpha k^\mu k^\nu(-3c_1+20c_3-48c_5) \nonumber \\
            & \hspace{1.8cm}  +a^\alpha k_\alpha g^{\mu\nu}[k^2(-9c_1+32c_3-48c_5)+4(c_2-c_4)]\} \nonumber \\
		\Pi_{4,a}^{\mu\nu} &= 16e^2 \{ 2a^\alpha k_\alpha k^\mu k^\nu(-c_1+12c_3-48c_5)
     \\
    & \ +a^\alpha k_\alpha g^{\mu\nu}[k^2(-5c_1+24_c3-48c_5)-4m^2(-c_1+3c_3)+2c_2-4c_4] \nonumber \\
    & \ +(a^\mu k^\nu+a^\nu k^\mu)[k^2(-c_1+12c_3-48c_5)+2m^2(c_1-6c_3)+c_2-4c_4]   \}. \nonumber
	\end{align}

    These results should conclude the analysis for the divergent part of \eqref{Seff}, but one can note the surface terms ($c_i$'s), which impregnate all results. This means further discussion is necessary to narrow down what one can obtain from one-loop computations for this theory. Furthermore, as expected, the only terms that can contribute to the generation of CFJ terms are the ones containing the tensors\footnote{The terms with the tensor $b^{(5)\mu\alpha\beta}_F$ could also contribute to the generation of the CFJ term, but as we computed, they are unambiguously zero.} $a^{(5)\mu\alpha\beta}_F$ and  $b^{(5)\mu\alpha\beta}$. The first is due to its indices symmetries \eqref{aF_exp}, and the latter is due to the nature of $\gamma_5$ traces. Although the results containing the tensor $a^{(5)\mu\alpha\beta}$ do not contribute directly to the generation of CFJ terms, they will play a role in constraining the relation between the surface terms, especially when we take the symmetry constraints for the theory into account.

	%\noindent $\bullet$\textbf{Terms with $b_F^{(5)\mu\alpha\beta}$}
	\subsection{Symmetry constraints - Ward–Takahashi identity} \label{secWardId}

    As noted in \cite{Jackiw2000}, in theories with finite radiative corrections, arbitrary parameters resulting from the cancellation of divergent integrals should be determined by the symmetries of the underlying model or its phenomenology. In a similar fashion, as we did in the past \cite{GazzolaEPL2013,Gazzola2012}, if one is dealing with a theory without anomalies, one can check if the one-loop corrections to the photon self-energy are transversal, \textit{i.e.}, one can expect that the Ward-Takahashi identity holds for such a theory. We then proceed to contract the external momentum with the whole amplitude $\sum_i\Pi_i$.

    It is clear that most of the terms are exactly zero, as the antisymmetric Levi-Civita tensor is contracted with the symmetric product $k_\alpha k_\beta$. Taking into consideration only the non-null contributions, this happens for the contraction with $\Pi_{4,b}$, $\Pi_{5,a_F}$, and $\Pi_{6,a_F}$. This means that if one constructs an extension for QED using the terms in the Lagrangian with the tensors $b^{(5)\mu\alpha\beta}$, $a^{(5)\mu\alpha\beta}_F$, and $b^{(5)\mu\alpha\beta}_F$, one could not fix on symmetry grounds the ambiguities emerging from the one-loop divergences of the theory, nor find any relations between the surface terms as in \cite{Gazzola2012}.

    As we mentioned earlier, this is where the results containing the tensor $a$ play their contribution to the analysis for the CFJ generation. Since its results are not antisymmetric to its indices $\mu,\nu$, the contraction with an external momentum will not yield zero term by term, and we can use this fact to find relations between the surface terms or fix some of them on symmetry grounds.
    
    This means that if one uses the self-energy transversality, \textit{i.e.},
    \begin{equation}
        k_\mu \sum_{n=1}^6\Pi_n^{\mu\nu} = k_\mu (\Pi_{2,a}^{\mu\nu}+\Pi_{3,a}^{\mu\nu}+\Pi_{4,a}^{\mu\nu})=0, \label{ward}
    \end{equation}
    one may find the relations between the surface terms
    \begin{subequations}
    \begin{align}
        -5c_1+33c_3-96c_5 &=0 \\
        -c_1+21c_3-96c_5 &= 0 \\
        -4m^2c_1+9c_2+24m^2c_3-24c_4 &=0 \\
        -12m^2c_1+5c_2+48m^2c_3-24c_4 &= 0 \\
        -4m^2c_1+5c_2+24m^2c_3-24c_4 &=0 \ .
    \end{align}
    \end{subequations}
    This system of equations only fixes
    \begin{equation}
        c_2 = 0 \ .
    \end{equation}
    The rest of the surface terms can be written in relation to $c_1$ as
    \begin{align}
        c_3 &= \frac{c_1}{3} \\
        c_4 &= m^2\frac{c_1}{6} \\
        c_5 &= \frac{c_1}{16}.
    \end{align}

        While the condition \eqref{ward} is expected to restrict the form of the allowed radiative corrections, in our analysis, we found that the Ward-Takahashi identity alone is not sufficient to fully fix the surface terms. Even after imposing gauge invariance through \eqref{ward}, some parameters, namely $c_1$, $c_3$, $c_4$, and $c_5$, remained undetermined. This outcome echoes the results obtained in \cite{Gazzola2012}, where the authors also found residual ambiguities in a model that included both minimal and non-minimal couplings.

        This reinforces an important aspect of IR: while it preserves gauge invariance at the algebraic level and makes the origin of ambiguities explicit, it does not automatically eliminate all regularization dependence. Instead, it isolates the ambiguity in well-defined terms such as \eqref{surface_term}, which can then be analysed and potentially fixed using additional physical or symmetry principles.

        In the context of the model under consideration \eqref{Lagrangiana}, the persistence of surface terms after applying the Ward-Takahashi identity identity suggests that the generation of the CFJ term is not completely universal, and still depends on regularization-induced parameters. As discussed in \cite{PetrovEPJC2021}, different regularization schemes yield different results for the CFJ coefficient, which is consistent with our findings.

        Thus, while IR offers a powerful tool to clarify the structure of divergences and their impact on Lorentz-violating terms, a complete determination of the induced CFJ term requires additional input — such as symmetry arguments beyond gauge invariance, renormalization conditions, or phenomenological constraints, as done in related works like \cite{GazzolaEPL2013}. %One possible candidate for such additional symmetry is the momentum routing invariance in the loops of the Feynman diagrams. This discussion was explored in the IR context by \cite{Ferreira2012}, and it should open up a venue for further investigation.
        A promising candidate for the additional symmetry required to constrain the remaining ambiguity is the invariance under momentum routing in the internal lines of Feynman diagrams. This aspect was systematically investigated within the Implicit Regularization framework in \cite{Ferreira2012}, providing a compelling direction for further theoretical exploration.

        This result contributes to the broader understanding that in Lorentz-violating theories, radiative corrections to CPT operators can carry inherent ambiguities, and their physical relevance must be evaluated case by case, depending on the regularization prescription and the set of imposed constraints.

    \subsection{On the generation of CFJ term}
   
    Collecting only the three amplitudes that yield a Levi-Civita structure, namely $\Pi_{4,b}^{\mu\nu}$, $\Pi_{5,a_F}^{\mu\nu}$ and $\Pi_{6,a_F}^{\mu\nu}$, we define
\[
\Pi_{\rm CFJ}^{\mu\nu}(k)
\;=\;
\Pi_{4,b}^{\mu\nu}(k)\;+\;\Pi_{5,a_F}^{\mu\nu}(k)\;+\;\Pi_{6,a_F}^{\mu\nu}(k)\,.
\]
After imposing the results for applying Ward–Takahashi identities, as relations among the $c_i$ (Section \ref{secWardId}), this can be written as

\begin{align}
		\Pi^{\mu\nu}_{\text{CFJ}} &= 
		\frac{8 i e^2}{3} k^\beta  \varepsilon^{\mu\nu}_{\ \  \alpha\beta}
		\left\{ c_1 [8k^2 a_F^\alpha+ (18 m^2-k^2) b^\alpha]
		 \nonumber \right. \\
		&\ \hspace{2.1cm} \left. -I_{\log}(m^2)
		\left[ (k^2+9m^2 )b^\alpha -2 k^2 a_F^\alpha
		\right]
		+9 I_{\rm quad}(m^2) b^\alpha
		\right\}\label{CFJ-full}
        %original sem relação entre c_i's
        %&= 
		%8 i e^2 k^\beta  \varepsilon^{\mu\nu}_{\ \  \alpha\beta}
		%\left\{ c_1 [8k^2 a_F^\alpha+ (12m^2-5k^2) b^\alpha]
		%+ c_2 [-4 a_F^\alpha + 3b^\alpha ] \nonumber \right. \\
		%& \left. \hspace{3cm} + c_3 k^2 [-16 a_F^\alpha + 86 b^\alpha]
		%- 36c_4 b^\alpha
		%-384k^2c_5 b^\alpha  \right\} \nonumber \\
		%&\ \ -8 i e^2 k^\beta \varepsilon^{\mu\nu}_{\ \  \alpha\beta}
		%\left\{ \frac{1}{3}I_{\log}(m^2)
		%\left[ (k^2+9m^2 )b^\alpha +2 k^2 a_F^\alpha
		%\right]
		%+3 I_{\rm quad}(m^2) b^\alpha
		%\right\}
	\end{align}
where $I_{\log}$ and $I_{\rm quad}$ denote intrinsic logarithmic \eqref{defIlog} and quadratic \eqref{defIquad} divergences, respectively, and $c_1$ is the remaining surface term, defined in \eqref{surface_term}.

From this result, two possible interesting lines of argument arise that deserve our attention. 

Firstly, as mentioned in the last subsection, the Implicit Regularization framework makes manifest that even after enforcing transversality, one surface term, $c_1$, remains undetermined.  Consequently, the CFJ coefficient itself carries an irreducible ambiguity proportional to $c_1$.  Physically, this reflects the fact that different regularization prescriptions, beyond DR, will shift the balance among $I_{\log}$, $I_{\rm quad}$, and $c_1$, altering the predicted CFJ coupling.  Thus, the induced CPT-odd operator is not entirely universal in perturbation theory: one must specify a renormalization or symmetry criterion to fix $c_1$ (cf.~\cite{GazzolaEPL2013}).

Secondly, one should compute our results using DR to construct a bridge between our results and those mentioned in \cite{PetrovEPJC2021}. If one evaluates $I_{\log}$, $I_{\rm quad}$ and all $c_i$ in DR, the distinction between logarithmic and quadratic divergences is lost (both appear as poles in $\epsilon$), and the remnant is exactly reduced to the CFJ coefficient reported in \cite{PetrovEPJC2021}.  More specifically, if one evaluates the mentioned terms explicitly using DR (with dimension $D=4-\epsilon)$, one finds
\begin{align}
I_{\rm quad}(m^2) &= \frac{i}{(4\pi)^2}\left( \frac{2}{\epsilon}-\gamma+1 +\ln\left(\frac{4\pi}{m^2}\right)\right) m^2, \\
I_{\log}(m^2) &= \frac{i}{(4\pi)^2} \left(\frac{2}{\epsilon}-\gamma +\ln\left(\frac{4\pi}{m^2}\right)\right) ,
\\
c_1 &= 0.
\end{align}
where $\gamma \approx 0.5772$ is the Euler-Mascheroni constant. Applying these results in equation \eqref{CFJ-full}, we reproduce the same result found in \cite{PetrovEPJC2021} (see Eq.(35) therein)
\begin{equation}
\Pi_{\rm CFJ}^{\mu\nu}(k)
= -\frac{e^2k^2}{3\pi^2\epsilon}\left(2a_F^\alpha-b^\alpha \right)k^\beta\varepsilon^{\mu\nu}_{\ \  \alpha\beta}+\text{finite terms}
\end{equation}
as expected. It is reasonable to note that other results may be achieved if one uses different regularization procedures due to the presence of the surface term in \eqref{CFJ-full}, which carries its intrinsic ambiguity.

Furthermore, in \cite{PetrovEPJC2021}, the authors adopt an ansatz relating the non-minimal tensors $a^{(5)\mu\alpha\beta}_F$ and $b^{(5)\mu\alpha\beta}$ (namely $2a_F^\alpha = b^\alpha$) precisely to cancel logarithmic divergences at one loop.  Although algebraically consistent, such a choice mixes structures of $I_{\log}$ and $I_{\rm quad}$ that in our scheme are distinct.  Since $I_{\log}$ and $I_{\rm quad}$ have different physical origins and scale dependence, enforcing their coefficients to cancel against each other may obscure the true scheme dependence.  In practice, one cannot assume that eliminating the logarithmic pole also guarantees the absence of quadratic sensitivity.  Any tensor ansatz of this type must be viewed with caution: It may remove one divergence, but leaves the other untouched and shifts the ambiguity into the remaining surface term $c_1$.

%In summary, our Implicit Regularization analysis confirms a radiatively generated CFJ term, whose Dimensional Regularization limit agrees with known results, but also highlights its intrinsic regularization ambiguity and questions the universality of the tensor relation they propose.

	%%% revisar aqui pra baixo. talvez jogue tudo fora
	
    %After the one-loop calculation of the polarization tensor, we found a result similar to the literature\footnote{Indeed, evaluating $I_{\rm quad}(m^2)$, $I_{\log}(m^2)$ and the surface terms explicitly using Dimensional Regularization, we find the same results [XXX] of \cite{PetrovEPJC2021}}, however with the advantage of having the split between divergences and surface terms (regularization dependent). 
    
    %This is in agreement with \cite{PetrovEPJC2021}, which shows different results when they used different approaches to deal with the integrals of the calculation.
	
	%The results with an intrinsic logarithmic divergence ($I_{\log}(m^2)$) show the possibility to find a relation between the $a^{(5)\mu\alpha\beta}$ and $b^{(5)\mu\alpha\beta}$ tensors, such that there is no logarithmic divergence. However, the result also indicates that a quadratic divergence ($I_{\rm quad}(m^2)$) will always remain. 

	\section{Conclusions}\label{conclusion}

    In this paper, we have employed the IR framework to analyse the one‐loop generation of Carroll–Field–Jackiw (CFJ) terms in a five‐dimensional CPT‐odd extension of quantum electrodynamics. By explicitly separating intrinsic divergences encoded in the logarithmic integral \(I_{\log}\) and the quadratic integral \(I_{\rm quad}\) from regularization dependent surface terms \(c_i\), IR makes manifest the ambiguities often concealed in DR or Cutoff schemes. Upon imposing the Ward–Takahashi identity, we found that all but one surface term could be expressed in terms of a single undetermined constant \(c_1\), demonstrating that gauge invariance alone is insufficient to uniquely fix the radiatively induced CFJ coefficient. 

    Moreover, in the DR limit, where both \(I_{\log}\) and \(I_{\rm quad}\) collapse into simple poles and all $c_i$ are null, our general expression reduces exactly to the CFJ term obtained by Petrov \textit{et al.} \cite{PetrovEPJC2021}. This agreement confirms the consistency of IR with established results while clarifying the hidden regularization assumptions underlying them. We have also critically examined the tensor relation ansatz introduced in previous work to eliminate logarithmic divergences. Within IR, such an ansatz merely trades one divergence for another, leaving \(I_{\rm quad}\) and \(c_1\) untouched, and thus does not resolve the fundamental scheme dependence of the induced CFJ term. Consequently, any prescription that relates non‐minimal tensors (such as \(a_F^\mu\) and \(b^\mu\)) should be justified on physical grounds, by symmetry principles or renormalization conditions. % beyond simple divergence cancellation.

    Our findings highlight that in Lorentz‐violating extensions of QED, radiative corrections to CPT operators inherently carry regularization ambiguities, similar to \cite{Gazzola2012}. Resolving these ambiguities will require supplementary input, whether from additional discrete symmetries, explicit renormalization prescriptions, or phenomenological constraints. More broadly, the IR approach presented here offers a transparent and systematic method for cataloging and controlling such ambiguities. Future work might explore additional symmetry principles, such as momentum routing invariance (discussed in \cite{Ferreira2012} within the context of IR), which may offer a potential path to fix the remaining regularization dependent term $c_1$; higher‐loop corrections, where overlapping divergences could introduce new classes of surface terms; extend the analysis to other CPT‐odd and CPT‐even operators of arbitrary mass dimension within the Standard‐Model Extension; or connect the residual ambiguity to experimental limits on Lorentz and CPT violation. We anticipate that this study will stimulate further research into the interplay between regularization, gauge invariance, and Lorentz‐violating physics in quantum field theory.

    \section*{Acknowledgements}

    H. G. Fargnoli was partially supported by FAPEMIG  RED-00133-21.

	\bibliographystyle{plain}

\end{document}